\documentclass[aps,prx,twocolumn,floatfix,showpacs,superscriptaddress]{revtex4-2}

\usepackage[unicode=true,
bookmarks=true,bookmarksnumbered=false,bookmarksopen=false, breaklinks=false,pdfborder={0 0 1},backref=false,colorlinks=true]{hyperref}
\hypersetup{linkcolor=magenta, citecolor=blue,urlcolor=blue}   
 
\usepackage{amsfonts}
\usepackage{amsmath,bbm}
\usepackage{amssymb}
\usepackage{amsthm}
\usepackage{amsbsy}
\usepackage{bm}
\usepackage{comment}
\usepackage{dcolumn}
\usepackage{epstopdf}
\usepackage{graphicx}
\usepackage{graphbox}
\usepackage{enumerate}
\usepackage{latexsym}
\usepackage{multirow}
\usepackage{mathtools}
\usepackage{subfig}
\usepackage{tabularx}
\usepackage{times}
\usepackage{tikz}
\usepackage{verbatim}
\usepackage{appendix}
\usepackage{ragged2e}
\usepackage[normalem]{ulem}
\usepackage{amsthm}
\usepackage{physics}

\newtheorem{criterion}{Criterion}

\makeatletter   
\long\def\@makecaption#1#2{%
  \vskip\abovecaptionskip
  {\small
  \justifying
  \hyphenpenalty=50 
  \exhyphenpenalty=50
  \sloppy              
  \noindent #1. #2\par}
  \vskip\belowcaptionskip}
\makeatother
\makeatletter   

\def\be{\begin{equation}} \def\ee{\end{equation}}
\def\bea{\begin{eqnarray}} \def\eea{\end{eqnarray}}

\tikzset{every picture/.style={line width=0.75pt}}

\begin{document}
\title{Boundary-sensitive non-Hermiticity of Floquet Hamiltonian: spectral transition and scale-free localization}

\author{Bo Li}
\thanks{These authors contributed equally to this work.}
\affiliation{ 
MOE Key Laboratory for Nonequilibrium Synthesis and Modulation of Condensed Matter, Shaanxi Province Key Laboratory of Quantum Information and Quantum Optoelectronic Devices, School of Physics, Xi’an Jiaotong University, Xi’an 710049, China}

\author{He-Ran Wang}
\thanks{These authors contributed equally to this work.}
\affiliation{
Chen-Ning Yang Institute for Advanced Study, Tsinghua University, Beijing, 100084, China}

\author{Fei Song}
\affiliation{
Kavli Institute for Theoretical Sciences and School of Quantum, University of Chinese Academy of Sciences, 100190 Beijing, China}

\begin{abstract}
We report a novel mechanism of boundary-sensitive $\mathcal{PT}$
symmetry breaking in one-dimensional Floquet systems. By designing
a time-periodic driving protocol, we realize a Floquet Hamiltonian
that is Hermitian under periodic boundary conditions yet acquires
non-Hermitian boundary terms under open boundary conditions due to
the noncommutativity of the driving Hamiltonians.
We show that boundary-induced $\mathcal{PT}$ breaking can occur when
the boundary term couples Bloch modes that become degenerate after their
quasienergy bands cross the frequency Brillouin zone boundary. For a single band,
this requires the quasienergy bandwidth to span the entire frequency
Brillouin zone, whereas in multi-band systems the relevant threshold
is determined by the bands that can be coupled by the boundary terms. In the $\mathcal{PT}$-broken phase, the eigenstates exhibit scale-free
localization arising from the system-size scaling of the induced
non-Hermitian terms. The associated wave-packet accumulation near the boundaries provides
a dynamical signature of the transition.
Finally, we present a general framework for constructing single- and
multiband models exhibiting this phenomenon.
\end{abstract} 

\maketitle

\section{Introduction}

In recent years, the discovery of non-Hermitian skin effect (NHSE) has sparked a surge of interest in non-Hermitian physics~\cite{yao2018edge,yao2018chern,kunst2018biorthogonal, lee2018anatomy,alvarez2017,Helbig2019NHSE,xiao2020non,Ghatak2019NHSE,Ghatak2019NHSE,Bergholtz2021RMP,Wang2022morphing, Ashida2021Non, Wang2024Tutorial, Gohsrich2024Perspective}. 
NHSE manifests as the exponential localization of eigenstates at the system boundaries under open boundary conditions (OBC), and the accompanying extreme sensitivity of the spectrum to boundary conditions, i.e., the OBC spectrum differs significantly from that under periodic boundary conditions (PBC). These findings challenge the conventional Bloch band theory, which relies on the invariance of the bulk band structure with respect to boundary conditions in the thermodynamic limit. To accurately characterize the localization of eigenstates and the OBC spectrum, the framework of non-Bloch band theory has recently been established~\cite{yao2018edge,yokomizo2020non,Longhi2019Probing,Song2019real,Kawabata2020nonBloch,Longhi2020chiral, Yang2020Auxiliary, Lee2020Unraveling, Yi2020, Xue2021simple, Hu2024Geometric, Li2024Classical, Wang2026NHselfenergy}. Central to this framework is the extension of the momenta associated with eigenstates from real to complex values.

Another cornerstone in the field of non-Hermitian physics is the exploration of parity-time ($\mathcal{PT}$) symmetric systems. In these systems, a transition from a real to complex spectrum, induced by varying system parameters, signifies $\mathcal{PT}$ symmetry breaking~\cite{PhysRevLett.80.5243, Bender_2007,Ozdemir2019, El-Ganainy2018, doi:10.1126/science.aar7709}. 
This transition is typically accompanied by the emergence of a defective eigenspace, where two or more eigenstates coalesce and fail to span the full Hilbert space. The corresponding transition point is referred to as an exceptional point (EP).
The unique properties of $\mathcal{PT}$ symmetry transitions and EPs have enabled a host of remarkable applications, including unidirectional invisibility~\cite{PhysRevLett.106.213901,Feng2013} and enhanced sensitivity~\cite{Hodaei2017, PhysRevLett.117.110802,Chen2018}. 
Recently, a novel mechanism for $\mathcal{PT}$ symmetry breaking has been identified that is driven solely by the NHSE and occurs only under OBC~\cite{Hu2024Geometric}.

In this work, we uncover a novel mechanism of boundary-sensitive $\mathcal{PT}$ symmetry breaking that is distinct from NHSE. We consider non-Hermitian one-dimensional (1D) Floquet systems driven by a time-periodic Hamiltonian with period $T$. We design the driving protocol such that the Floquet Hamiltonian is Hermitian and exhibits a purely real quasienergy spectrum under PBC. In stark contrast, under OBC, the non-commutativity of the driving Hamiltonians at different times generates $\mathcal{PT}$ symmetric non-Hermitian boundary terms in the Floquet Hamiltonian. These boundary terms can induce $\mathcal{PT}$ symmetry transitions when the parameters exceed a threshold.
Remarkably, we establish that this threshold is related to the quasienergy spectrum winding.

Furthermore, in the $\mathcal{PT}$-broken phase where the quasienergy spectrum develops nonzero imaginary parts, we identify the phenomenon of scale-free localization, characterized by the $1/N$ scaling of the imaginary part of both the complex momenta and the eigenvalues, where $N$ is the system size.
The term ``scale-free localization'' stems from the invariance of the eigenstate wavefunctions under length-scale transformations.
Recently, such localization behaviors have been reported in static non-Hermitian systems, such as coupled chains with opposite localization directions of NHSE~\cite{Li2020, PhysRevB.104.165117}, systems with disordered hopping across the boundaries~\cite{Li2021,Molignini2023Anomalous,Zheng2024Exact}, and those with single
non-Hermitian impurity~\cite{Li2023Scale, Wang2023Scale}. By contrast, the mechanism uncovered in this work relies on the Floquet protocol, where the effective non-Hermitian impurities naturally arise from higher-order terms beyond the time-averaged Hamiltonian.

We note that non-Hermitian Floquet systems have also attracted considerable attention because of their rich topological properties, including anomalous Floquet phases and boundary modes~\cite{Zhou2018Non, Wu2020Floquet, Zhou2021Dual, Liu2022Symmetry, Lin2024Observation, Zhou2026Topology}. In contrast to these studies, which primarily focus on the emergence and characterization of boundary states under OBC, the phenomenon uncovered by this work is the boundary sensitivity of the bulk spectrum.

This paper is organized as follows. In Sec.~\ref{sec:simple_model}, we introduce a minimal model to illustrate the construction of $\mathcal{PT}$ symmetric Floquet systems. In Sec.~\ref{sec:PTbreaking}, we demonstrate the critical role of boundary conditions in the $\mathcal{PT}$ symmetry breaking, followed by an analysis of the emergence of scale-free localization in Sec.~\ref{sec:scale_free}. Generalizing beyond the minimal model, in Sec.~\ref{sec:condition} we outline a recipe for constructing broader classes of Floquet systems that exhibit similar properties. Sec.~\ref{sec:signature} is devoted to discussing the $\mathcal{PT}$ breaking criterion for general cases.
Applying this framework, we extend our discussion to multi-band systems and further clarify the conditions for $\mathcal{PT}$-symmetry breaking in Sec.~\ref{sec:multi-band}.  In Sec.~\ref{sec:dynamics}, we study the dynamical signature of PT breaking under OBC. Finally, we conclude with a summary and outlook in Sec.~\ref{sec:conclusion}. Technical derivations and details are provided in the Appendices.

\section{Non-Hermitian Floquet system}\label{sec:simple_model}

We illustrate our approach using a time-periodic Hamiltonian $H(\tau) = H(\tau+T)$, which consists of two stages of non-unitary evolution:
\begin{equation}\label{eq:Hamiltonian}
  H(\tau)=\begin{cases}
    H_1 = t\hat{L}, & 0\leq \tau< T/2.\\
    H_2 =t\hat{R}, & T/2\leq \tau< T,
  \end{cases}
\end{equation}
with
\begin{align}\label{eq:simple_model}
&\hat L=\sum_{j=1}^{N-1} |j\rangle\langle j+1| +\eta|N\rangle\langle 1|,\nonumber\\
&\hat R=\sum_{j=1}^{N-1} |j+1\rangle\langle j| + \eta|1\rangle\langle N|.
\end{align}
Here $\tau$ denotes the time, $t$ is the hopping amplitude, $\hat L$, $\hat R$ denote the left and right one-site shift operators on an $N$-site 1D chain, respectively. 
Boundary conditions are controlled by the parameter $\eta$, with $\eta=1$ for PBC and $\eta = 0$ for OBC.

The dynamics of the system are characterized by the stroboscopic Floquet operator:
\begin{eqnarray}\label{eq:Floquetoperator}
U_F=\mathcal T e^{-i\int_0^T d\tau H(\tau)}=e^{-iH_2T/2}e^{-iH_1T/2} = e^{-i H_FT},
\end{eqnarray}
where $\mathcal T$ denotes the time ordering. Properties of the system are encoded in the effective Floquet Hamiltonian $H_F$, of which the quasienergy spectrum and eigenstates are the primary focus of our work. 
We set the branch cut at $\pm\pi/T$, restricting the real part of the quasienergy spectrum to the frequency Brillouin zone $[-\pi/T,\pi/T]$. 

Next, we investigate the properties of $H_F$ under different boundary conditions.
Under PBC, the shift operators commute due to translation symmetry and can be simultaneously diagonalized in the momentum basis. This leads to a Hermitian Floquet Hamiltonian:
\begin{eqnarray}\label{eq:PBCHamiltonian}
H_{F,\text{PBC}}=\frac{\lambda}{T}(\hat L + \hat R) = \frac{2\lambda}{T}\sum_k \cos k|k\rangle\langle k|, 
\end{eqnarray}
where $\lambda=tT/2$ is a dimensionless parameter, $k=2\pi n/N$ ($n=0,1,2,\cdots, N-1$) is the crystal momentum, and $\ket{k}$ denotes the plane-wave state of momentum $k$. The quasienergy spectrum under PBC is purely real due to Hermiticity.

On the other hand, under OBC, translational symmetry is broken, and the operators $\hat{L}$ and $\hat{R}$ no longer commute, leaving nonzero commutators with support near the boundaries. 
Consequently, the effective Hamiltonian $H_{F,\text{OBC}}$ is not simply the sum of $H_1$ and $H_2$. As we will demonstrate, the higher-order terms arising from this non-commutativity make the Floquet Hamiltonian non-Hermitian, leading to rich phenomena including $\mathcal{PT}$ symmetry breaking and scale-free localization.

\begin{figure}
\hspace*{-0.5\textwidth}
\begin{tabular}{cc}
\includegraphics[width=1\linewidth]{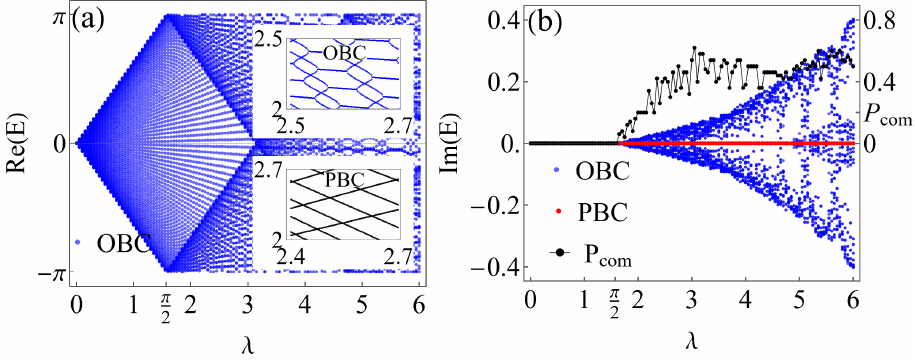} \\
\includegraphics[width=1\linewidth]{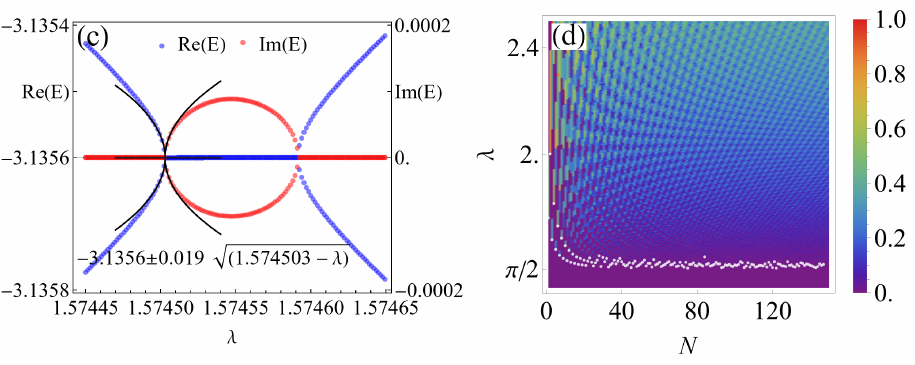} 
\end{tabular}
\caption{(a) Real part of the quasienergy spectrum under open boundary conditions, as a function of $\lambda$. Inset shows the zoomed-in spectrum under open and periodic boundary conditions for model~\eqref{eq:Hamiltonian}. (b) Imaginary part of the quasienergy spectrum under open boundary conditions as a function of $\lambda$. The proportion of complex eigenvalues $P_{\text{com}}$ is superimposed (scale on the right axis). (c) Magnified view of the spectrum in the vicinity of an exceptional point. Solid black lines represent a square-root fit characterizing the bifurcation of the imaginary part. (d) Phase diagram created using $P_{\text{com}}$ as the order parameter, where white dots represent the threshold value of $\lambda$ for $\mathcal{PT}$ symmetry breaking, as a function of system size $N$. We set $T=1$.}\label{fig:spectrum}
\end{figure}

We first show the $\mathcal{PT}$ symmetry of $H_F$. The parity ($\hat P$) and time-reversal ($\hat K$) operators are defined as:
\begin{eqnarray}
\hat P \ket{j} =\ket{N-j+1},\qquad
\hat K i\hat K^{-1}=-i.
\end{eqnarray}
It follows that $\hat P H_1 \hat P^{-1}=H_2$, and therefore
\begin{equation}
\hat P \hat K U_F \hat K^{-1}\hat P^{-1}=U_F^{-1},
\end{equation}
and $\hat P \hat K H_F \hat K^{-1}\hat P^{-1}=H_F$. 

In Fig.~\ref{fig:spectrum}(a, b), we plot the real and imaginary parts of quasienergies as a function of $\lambda$ for $T=1$. For the remainder of this paper, we set $T=1$ for convenience. 
The inset of Fig.~\ref{fig:spectrum}(a) shows signatures of numerous exceptional points (EPs) in the OBC spectrum. In Fig.~\ref{fig:spectrum}(b), as $\lambda$ varies, we superimpose the proportion of complex eigenvalues $P_{\text{com}} = n_{\text{com}}/N$ onto the imaginary part of the spectrum, where $n_{\text{com}}$ is the number of complex eigenvalues.
The onset of the $\mathcal{PT}$ symmetry breaking is observed for $\lambda$ larger than $\lambda_c=\pi/2$.
In Fig.~\ref{fig:spectrum}(c), we focus on the vicinity of an EP formed by a pair of eigenvalues, which clearly demonstrates the coalescence of the real parts accompanied by the bifurcation of the imaginary parts. By fitting the imaginary part of the quasienergy $\text{Im}(E)$ against $\lambda$ with a square-root function, we confirm that the EP is of second order.
Finally, in Fig.~\ref{fig:spectrum}(d), we analyze the finite-size scaling of the threshold using $P_{\text{com}}$ as an order parameter. We find that the critical value $\lambda_c$ asymptotically approaches $\pi/2$ as the system size $N$ increases.

\section{$\mathcal{PT}$ symmetry breaking under OBC}\label{sec:PTbreaking}

In this section, we explore the mechanism underlying $\mathcal{PT}$ symmetry breaking under OBC.
We begin by expressing the OBC Floquet Hamiltonian in Eq.~\eqref{eq:Floquetoperator} formally as a series in the dimensionless parameter $\lambda$ using the Baker-Campbell-Hausdorff (BCH) formula~\cite{vajna2018replica}:
\begin{equation}\label{eq:BCHexpansion}
H_{F,\text{OBC}} T=\lambda (\hat L+\hat R)-\frac{i\lambda^2}{2}[\hat L,\hat R]+ \frac{\lambda^3}{12}[\hat L-\hat R,[\hat L,\hat R]] + \cdots.
\end{equation}
The series contains infinitely many terms of increasing order in $\lambda$. 
The leading-order term in $\lambda$ reads $H_0=\frac{\lambda}{T}(\hat L+\hat R)$, which amounts to the averaged Hamiltonian. In the thermodynamic limit, $H_0$ shares the same purely real quasienergy spectrum as the PBC Hamiltonian. The higher-order corrections, defined as $V \equiv H_{F,\text{OBC}}-H_0$, break both translational symmetry and Hermiticity.
For example, the second-order term explicitly introduces non-Hermitian boundary terms:
\begin{equation}
    -\frac{i\lambda^2}{2}[\hat L,\hat R] = -\frac{i\lambda^2}{2}(\ket{1}\bra{1}-\ket{N}\bra{N}).
\end{equation}
Crucially, the magnitude of the matrix elements in $V$ decays exponentially from the boundaries into the bulk, as shown numerically in 
Appendix~\ref{appendix:perturbation}. We can therefore treat $V$ as a $\mathcal{PT}$ symmetric perturbation to $H_0$.

From the perspective of perturbation theory, second-order EPs can emerge when a pair of energy levels of the unperturbed Hermitian part approach each other, allowing non-Hermitian perturbations to induce level attraction and the coalescence of the associated eigenstates into a defective two-level subspace. In our model, this can never happen for $\lambda < \lambda_c = \frac{\pi}{2}$, as the quasienergy spectrum of $H_0$ is confined to the interval $[-2\lambda/T, +2\lambda/T]$ and exhibits no level crossings, which is supported by the absence of EPs in this regime, as shown in Fig.~\ref{fig:spectrum}(a,b).

On the other hand, when $\lambda \ge \lambda_c$, the energy levels at the band edges (top and bottom) exceed the boundaries of the frequency Brillouin zone $[-\pi/T, \pi/T]$. Due to the periodicity of the quasienergy, these levels are folded back into the Brillouin zone by taking the quasienergy modulo $2\pi/T$, resulting in level crossings as depicted in the inset of Fig.~\ref{fig:spectrum}(a) for the PBC spectrum (equivalent to the spectrum of $H_0$ for large $N$). These crossings provide the necessary degeneracy for the non-Hermitian perturbation $V$ to generate EPs in the OBC spectrum.

\begin{figure}
\hspace*{-0.5\textwidth}
    \includegraphics[width=1\linewidth]{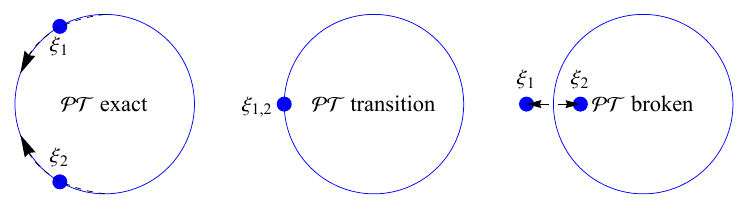}
 \caption{Illustration of $\mathcal{PT}$ symmetry transition by tracking trajectories of a pair of eigenvalues $\xi_{1,2}$ of the Floquet operator $U_F$ under open boundary conditions.}\label{fig:trajectory}
\end{figure}

This mechanism is illustrated in Fig.~\ref{fig:trajectory} via the trajectories of the eigenvalues of the Floquet operator $U_F$. As shown, two eigenvalues $\xi_{1,2}$ approach each other along the unit circle in the complex plane as $\lambda$ increases, eventually colliding at the critical value.  Upon $\mathcal{PT}$-symmetry breaking, the eigenvalues depart from the unit circle, developing non-unit moduli while satisfying the condition $|\xi_1 \xi_2|=1$ (with $|\xi_1|, |\xi_2| \neq 1$).

Based on the observations above, we propose the following criterion for the $\mathcal{PT}$ symmetry breaking:
\begin{criterion}\label{criterion_1}
\textit{In the thermodynamic limit, for a single-band model}, the quasienergy spectrum of $H_{F,\mathrm{OBC}}$ begins to develop exceptional points when the bandwidth of $H_{F,\mathrm{PBC}}$ reaches $2\pi/T$, spanning the entire frequency Brillouin zone.
\end{criterion}
\noindent In Sec.\ref{sec:signature}, we derive, through a rigorous analysis, a more general criterion applicable to multi-band models. Criterion~\ref{criterion_1} is recovered as its single-band limit.


We remark on the distinction between the mechanism of $\mathcal{PT}$ breaking in our Floquet system and that in static models.
In static non-Hermitian systems, the coalescence of energy levels into a second-order EP typically occurs between two distinct bands, where increasing non-Hermiticity induces a band touching~\cite{Heiss_2012,El-Ganainy2018, doi:10.1126/science.aar7709}. In contrast, our minimal construction in Eq.~\eqref{eq:Hamiltonian} requires only a single band under PBC. Here, the periodicity of the frequency Brillouin zone allows the single band to intersect with itself across the Brillouin zone boundaries, thereby enabling the formation of EPs under OBC. This mechanism is conceptually analogous to the formation of topologically protected Floquet edge modes at quasienergies $0$ and $\pi$ in Hermitian systems with zero Chern number~\cite{kitagawa2010topological,jiang2011majorana,rudner2013anomalous}.

\section{Scale-free localization}\label{sec:scale_free}

In this section, we investigate the order of non-Hermitian perturbations to quasienergies and eigenstates.
In the BCH formula Eq.~\eqref{eq:BCHexpansion}, for moderate energy scales, i.e., $\lambda\sim O(1)$, the higher-order terms $V$ generated by nested commutators are strongly localized at boundaries, which holds even for parameters beyond the convergence radius $\lambda_*$ (see Appendix~\ref{appendix:perturbation}). Thus, the perturbation can be divided into a boundary term $V_{\text{boundary}}$ and a bulk term $V_\text{bulk}$. We will demonstrate that both terms contribute corrections of order $O(1/N)$ to eigenstates and eigenvalues. 
The boundary term $V_{\text{boundary}}$ contains sizable and approximately size-independent elements, which should influence bulk eigenvalues and eigenstates at order $\sim 1/N$~\cite{Li2023Scale}: intuitively, in the basis of $H_0$ eigenstates $|\psi_n\rangle$ with $H_0|\psi_n\rangle=\varepsilon^0_n|\psi_n\rangle$, the perturbative matrix elements read 
\begin{equation}
   \langle\psi_m|V_{\text{boundary}}|\psi_n\rangle\sim O(1)\times(N^{-1/2})^2\sim N^{-1}.
\end{equation}
The bulk term $V_{\text{bulk}}$, on the other hand, has amplitudes that scale as $\sim 1/N$, as evidenced by numerical results (see Appendix~\ref{appendix:perturbation}).

The scaling analysis of the perturbative matrix elements suggests that in the $\mathcal{PT}$-broken phase, the imaginary part of quasienergies scales as $\mathcal{O}(1/N)$. To characterize the spatial structure of eigenstates, we adopt the standard standing-wave ansatz for the wavefunction:  $|\psi_N(\beta)\rangle=\sum\limits_{j=1}^N (A_+ \beta^j+A_- \beta^{-j})|j\rangle$. The associated quasienergy is determined by the dispersion relation 
\begin{equation}\label{eq:imaginary_energy}
    E = \frac{\lambda}{T}(\beta + \beta^{-1}).
\end{equation}
In the Hermitian case, $\beta = e^{ik}$ lies on the complex unit circle, where $k$ represents the crystal momentum. 
By adding non-Hermitian boundary perturbations, $\beta$ can deviate from the unit circle. Given the $\mathcal{O}(1/N)$ scaling of the imaginary energy component, it should scale as $|\beta| \approx 1+\alpha/N$. 
This scaling leads to the scale-free localization of the wavefunction profile $|\psi_N(x)|\simeq |\beta|^x \simeq e^{\alpha x/N}$, which respects the scale invariance $|\psi_N(x)|\simeq |\psi_{sN}(sx)|$ for a scaling factor $s$.

\begin{figure}
\hspace*{-0.5\textwidth}
    \includegraphics[width=1\linewidth]{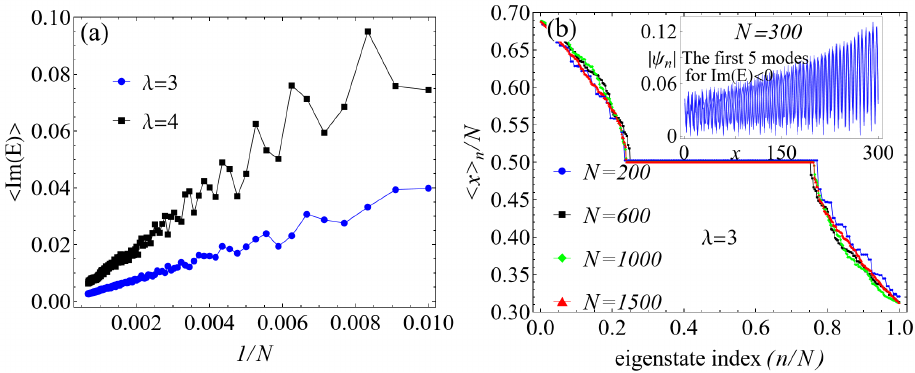}
 \caption{Scale-free localization in the $\mathcal{PT}$-broken phase for model~\eqref{eq:Hamiltonian}. (a) Finite-size scaling of the averaged imaginary part of quasienergies. $\lambda=3$ and $4$. (b) The eigenstate-resolved mean position ratio $\langle x\rangle_n/N$, plotted for various system sizes $N$. The eigenstate index $n$ is sorted in ascending order of the imaginary part of the corresponding quasienergy. The inset displays wavefunction profiles of selected eigenstates.
}\label{fig:scale_free}
\end{figure}

In Fig.~\ref{fig:scale_free}(a), we perform a finite-size scaling of the mean imaginary part of quasienergies. The results explicitly show $ 1/N$ scaling in the large-$N$ limit. To further corroborate the scale-free localization, we calculate the mean position of the $n$-th eigenstate, defined as $\langle x\rangle_n = \sum\limits_{j=1}^Nj|\bra{j}\psi_{n}\rangle|^2 \big/ \sum\limits_{j=1}^N|\bra{j}\psi_{n}\rangle|^2$. As shown in Fig.~\ref{fig:scale_free}(b), the mean position ratio $\langle x\rangle_n/N$ deviates from the center value ($0.5$) for a substantial fraction of the eigenstates. Remarkably, the fraction remains consistent across various system sizes $N$, confirming the scale-invariant feature of the wavefunction profiles.


\section{General constructions}\label{sec:condition}

To generalize the construction presented in Eq.~\eqref{eq:Hamiltonian}, we identify the following three essential requirements for the driving Hamiltonians $H_{1,2}$:
\begin{enumerate}
    \item[(i)]\emph{$\mathcal{PT}$ symmetry}. $\hat P \hat K H_{1(2)} \hat K^{-1}\hat P^{-1}=H_{2(1)}$,
    \item[(ii)]\emph{Hermiticity under PBC}. $(\hat H_1+\hat H_2)^\dagger=\hat H_1+\hat H_2$, and $[\hat H_1,\hat H_2]_{\text{PBC}}=0$;
    \item[(iii)]\emph{Non-Hermiticity under OBC}. $[\hat H_1,\hat H_2]_{\text{OBC}}\neq 0$.
\end{enumerate}
The first condition imposes $\mathcal{PT}$ symmetry for the Floquet operator $U_F$ and Floquet Hamiltonian $H_F$, where the parity operator $\hat P$ is model-dependent and not necessarily restricted to the spatial reflection operator. The second condition guarantees that $H_{F,\text{PBC}}$ is Hermitian, while the third one is the source of non-Hermitian terms in $H_{F,\text{OBC}}$. 

We propose a general ansatz for multi-band systems satisfying these conditions:
\begin{eqnarray}\label{eq:general_form}
 H_i=\mathbbm{1}\otimes\hat A_i+\sum_{r=1}^w [\hat L^r\otimes\hat X_i^{(r)} + \hat R^r\otimes\hat Y_i^{(r)}],\quad  (i=1,2)\nonumber\\
\end{eqnarray}
where $\hat A_i$ describes intracell couplings, and $\hat X^{(r)}_i$ and $\hat Y^{(r)}_i$ represent intercell hoppings between unit cells separated by a distance $r$.
Condition (ii) imposes the following constraints on the Bloch Hamiltonian:
\begin{eqnarray}
&&\hat h^\dagger_1(k)+\hat h^\dagger_2(k)=\hat h_1(k)+\hat h_2(k),\label{eq:cond1}\\
&&[\hat h_1(k),\hat h_2(k)]=0\label{eq:cond2},
\end{eqnarray}
where $\hat h_i(k)=\hat A_i+\sum\limits_{r=1}^w [e^{ikr}\hat X^{(r)}_i+e^{-ikr}\hat Y^{(r)}_i]$. Condition (iii) is more subtle, as it requires specific boundary terms to survive the BCH expansion. While the detailed derivation is provided in Appendix~\ref{appendix:commutator}, we find that at least one of the following conditions on the hopping matrices must be satisfied:
\begin{eqnarray}\label{eq:cond3}
&&[\hat X^{(r)}_1, \hat Y_2^{(r')}]\neq 0, \quad
[\hat Y^{(r)}_1, \hat X_2^{(r')}]\neq 0, \nonumber\\
&&\hat Y_2^{(r')}\hat X_1^{(r)}-\hat X_2^{(r')}\hat Y_1^{(r)}\neq 0,\quad  r,r'=1,2,\cdots w.
\end{eqnarray} 
Eqs.~\eqref{eq:general_form} to \eqref{eq:cond3} constitute a general recipe for constructing Floquet systems that exhibit boundary-induced $\mathcal{PT}$ symmetry breaking.

Based on the conditions above, the most straightforward construction satisfies 
\begin{eqnarray}\label{eq:step_sharp_relation}
H_2= H_1^\dagger.    
\end{eqnarray}
In the following section, we will focus on this situation. We first provide a general discussion on $\mathcal{PT}$ breaking criteria; then construct concrete multi-band models and demonstrate that the $\mathcal{PT}$ transition under OBC is intimately linked to the winding of the quasienergy spectrum under PBC.

\section{Sign characteristic and boundary-induced $\mathcal{PT}$ breaking}
\label{sec:signature}

In this section, we focus on the two-step driving protocol and establish a general $\mathcal{PT}$-breaking criterion for the case following Eq.~\eqref{eq:step_sharp_relation}.  We introduce the sign characteristic of the PBC Floquet
modes and show how it determines whether an OBC boundary perturbation can
produce an exceptional point (EP). Then, based on the theory, we summarize a general $\mathcal{PT}$ breaking criterion valid for both single- and multi-band models. 

\subsection{Sign characteristic of the PBC Floquet modes}

We consider a general two-step Floquet evolution
\begin{equation}
U_F
=
e^{-iH_2T/2}e^{-iH_1T/2},
\label{eq:two_step_floquet_signature}
\end{equation}
which obeys the relation~\eqref{eq:step_sharp_relation}. Under PBC, translation symmetry allows us to work at each momentum \(k\). We can work in the PBC eigenbasis: $U_{F}(k)|u_n(k)\rangle
=e^{-i\theta_n(k)}|u_n(k)\rangle$. By construction, the Floquet Hamiltonian
$H_F(k)=(i/T)\log U_F(k)$ is Hermitian under PBC,
so the left and right eigenstates of $U_F(k)$ can be
chosen as Hermitian conjugate pairs.  Using the form of Eq.~\eqref{eq:step_sharp_relation}, we can show the following relation:
\begin{equation}\label{eq:real_quantity}
e^{i\theta_n(k)/2}q_n(k)
=
\left[
e^{i\theta_n(k)/2}q_n(k)
\right]^*,
\end{equation}
where $q_n(k)
=
\langle u_n(k)|
e^{-iH_1(k)T/2}|u_n(k)\rangle$, so this quantity is real, see details in Appendix~\ref{app:signature}. We therefore define the PBC sign characteristic
as
\begin{equation}
\nu_n(k)
=
\operatorname{sgn}
\left[
e^{i\theta_n(k)/2}q_n(k)
\right].
\label{eq:PBC_signature_final}
\end{equation}
For a $\mathcal{PT}$-exact mode lying on the unit circle, \(\nu_n(k)=\pm1\).

In evaluating Eq.~\eqref{eq:PBC_signature_final}, \(\theta_n(k)\) should
be chosen as a continuous unwrapped eigenphase along the corresponding PBC
band, namely, $\theta_n(k)=E_n(k)+2\pi m$ where $E_n(k)$ is the quasi-energy in the principal logarithm branch, and $m\in\mathbbm{Z}$. If the same Floquet eigenvalue is represented on two neighboring
Floquet sheets: $\theta_n(k)\rightarrow\theta_n(k)+2\pi$, then $e^{i[\theta_n(k)+2\pi]/2}
=
-e^{i\theta_n(k)/2}$. Thus, two resonant branches related by an odd Floquet-zone shift naturally
carry opposite sign characteristics. This sheet parity is the origin of
the opposite-signature collision in the minimal single-band model.

\subsection{Two-mode projection and the opposite-signature condition}

Consider two PBC modes $\{|\psi_1\rangle, |\psi_2\rangle\}$ that become resonant modulo \(2\pi\). In their
two-dimensional subspace, choose a basis in which the signature metric is
\begin{equation}
\eta=
\begin{pmatrix}
\nu_1&0\\
0&\nu_2
\end{pmatrix}.
\label{eq:two_mode_signature_metric}
\end{equation}
Since both the PBC drive and the OBC boundary deformation preserve the Hermitian conjugate structure between $H_1$ and $H_2$, the leading effective operator in this
subspace can be written as
\begin{equation}
U_{\rm eff}
=
\xi_0
\exp\left(-\frac{i}{N}K\right)
+
O(N^{-2}),
\label{eq:Ueff_K_signature}
\end{equation}
where \(\xi_0\) is the resonant unit-circle eigenvalue (under PBC). Eq.~\eqref{eq:Ueff_K_signature} can be regarded as the OBC evolution operator in the space spanned by the two PBC resonance modes; the exponential term denotes the boundary correction induced coupling between two PBC modes. Based on the relation in Eq.~\eqref{eq:step_sharp_relation}, we obtain the following projected adjoint relation for $K$
\begin{equation}
K^\dagger\eta=\eta K.
\label{eq:K_pseudo_Hermitian}
\end{equation}
The detailed derivation is given in the Appendix~\ref{app:projected_constraint}.

We first consider two modes with the same signature, namely, the two modes fall into two Floquet sheet with both even or odd sheet number. In this case,
\(\eta=\pm \mathbbm{1}\), and Eq.~\eqref{eq:K_pseudo_Hermitian} reduces to
\begin{equation}
K^\dagger=K.
\end{equation}
Therefore, the eigenvalues of \(K\) are real. From
Eq.~\eqref{eq:Ueff_K_signature}, the corresponding Floquet eigenvalues
remain on the unit circle. Therefore, it cannot generically produce $\mathcal{PT}$ breaking.

For opposite signatures, we choose the signature matrix
\begin{equation}
\eta=
\begin{pmatrix}
1&0\\
0&-1
\end{pmatrix}.
\label{eq:indefinite_eta}
\end{equation}
Equation~\eqref{eq:K_pseudo_Hermitian} then gives
\begin{equation}
K=
\begin{pmatrix}
a&b\\
-b^*&d
\end{pmatrix},
\qquad
a,d\in\mathbb R.
\label{eq:opposite_signature_K}
\end{equation}
The two eigenvalues of \(K\) are $\kappa_\pm
=
\frac{a+d}{2}
\pm
\sqrt{
\frac{(a-d)^2}{4}
-
|b|^2
}$. They coalesce when $|b|=\frac{|a-d|}{2}$,
which is the EP condition in the projected subspace. For $|b|>\frac{|a-d|}{2}$,
the square root in the eigenvalues is imaginary and
\(\kappa_+\) and \(\kappa_-\) form a complex-conjugate pair. Consequently,
\begin{equation}
\xi_\pm
=
\xi_0e^{-i\kappa_\pm/N}
\end{equation}
leave the unit circle as a reciprocal-conjugate pair. This explicitly shows
why a $\mathcal{PT}$-breaking collision requires opposite signatures. At the EP, the coalesced state becomes null with respect to the norm,
\begin{equation}
\langle\psi_{\rm EP}|e^{-iH_1T/2}|\psi_{\rm EP}\rangle=0.
\label{eq:null_state_at_EP}
\end{equation}
Beyond the EP, the individual eigenvalues no longer lie on the unit circle,
and the sign characteristic in
Eq.~\eqref{eq:PBC_signature_final} is no longer separately defined for the two broken modes.

\subsection{General $\mathcal{PT}$ breaking criterion.}

It is now clear that, according to Criterion \ref{criterion_1}, only when the PBC bandwidth goes over $2\pi/T$ can the degenerate modes have opposite signatures, thus allowing $\mathcal{PT}$ breaking. Moreover, the PBC-degenerate modes considered above need not originate from the same band. The conclusion applies to arbitrary degenerate modes, whether from the same or different PBC bands, as long as boundary-induced coupling between them is allowed. We therefore arrive at the following general criterion for \(\mathcal{PT}\) breaking:
\begin{criterion}\label{criterion_2}
\textit{For two-step driving systems satisfying $H_2=H_1^\dagger$, in the thermodynamic limit, $\mathcal{PT}$ breaking in the quasienergy spectrum of $H_{F,\mathrm{OBC}}$ can occur if the following conditions are satisfied:
\begin{enumerate}
\item Boundary-induced coupling between the degenerate modes involved is allowed by symmetry;
\item The two modes have opposite sign characteristics $\nu(k)$; namely, they lie on different Floquet sheets with opposite parities.
\end{enumerate}
}
\end{criterion}
The second condition is equivalent to requiring that the total bandwidth spanned by the coupled bands reach at least \(2\pi/T\), provided that coupling between them is symmetry allowed. Otherwise, one can always choose an appropriate branch cut such that the two degenerate modes lie on the same Floquet sheet. It is evident that Criterion~\ref{criterion_2} encompasses Criterion~\ref{criterion_1}: for a single-band model, the first condition is automatically satisfied, whereas the second condition requires that \(\mathcal{PT}\) breaking emerge only when the bandwidth reaches \(2\pi/T\). In the following, we show that the \(\mathcal{PT}\)-breaking thresholds of multi-band models can also be predicted by this criterion.

\begin{figure}
\hspace*{-0.45\textwidth}
\begin{tabular}{cc}
\includegraphics[width=0.9\linewidth]{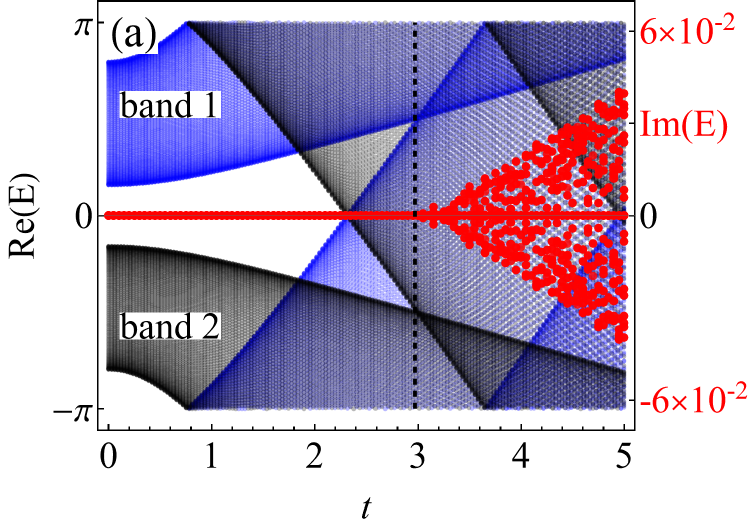}  \\
\includegraphics[width=0.9\linewidth]{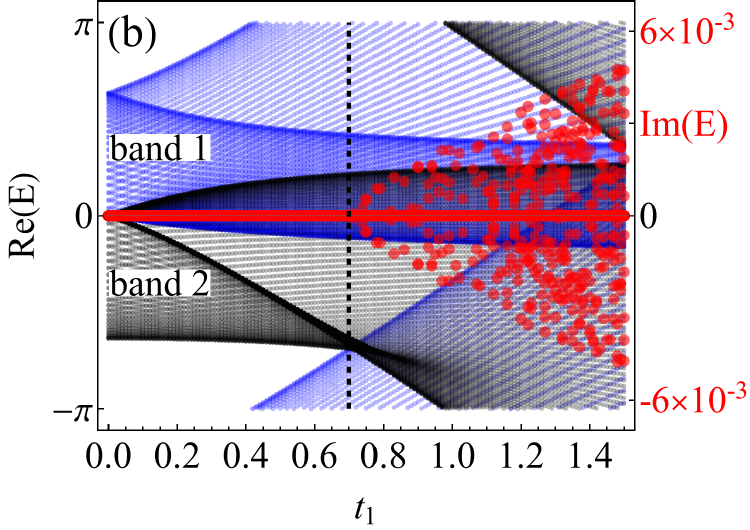} 
\end{tabular}  
\caption{Quasienergy spectra of the two-band models as functions of certain parameters. (a) Type-I model [Eq.~\eqref{eq:twoband1}], varying $t$. (b) Type-II model [Eq.~\eqref{eq:twoband2}], varying $t_1$ and fixing $t_2=0.5$. The vertical black dashed line marks the threshold for $\mathcal{PT}$ symmetry breaking. The real parts of quasienergies are represented by blue and black curves for two bands (scale on the left axis); the imaginary parts are shown in red (scale on the right axis). We set $T=1$.
 }\label{fig:twoband}
\end{figure}

\section{Multi-band models}\label{sec:multi-band}

In this section, we investigate two types of multi-band models: (I) all the intra-cell ($\hat A_i$) and inter-cell ($\hat X_i^{(r)}, \hat Y_i^{(r)}$) hopping matrices mutually commute, such that the Hamiltonian decomposes into independent single-band sectors; (II) a genuinely multi-band model where the hopping matrices do not commute. Here, boundary terms can induce inter-band mixing, leading to qualitatively different $\mathcal{PT}$ breaking thresholds.

We first consider a Type-I model defined by:
\begin{eqnarray}\label{eq:twoband1}
\hat{H}_{1(2)}^{(\text{I})}=\mathbbm{1}\otimes\frac{3}{2}(t\hat\sigma_x+\hat\sigma_z)+\hat O_{1(2)}\otimes (t\hat\sigma_x+\hat\sigma_z),
\end{eqnarray}
where $\hat O_{1(2)}=\hat L (\hat R)$. In this setup, the intracell and intercell couplings share the same matrix structure, ensuring that all hopping matrices commute. The boundary-induced perturbation cannot couple two bands, see Appendix~\ref{app:typeI}. The construction satisfies the three conditions with $\hat P$ acting as the spatial reflection. The quasienergy spectrum as a function of the hopping parameter $t$ is presented in Fig.~\ref{fig:twoband}(a). As expected for a decoupled system, the $\mathcal{PT}$-symmetry breaking threshold is reached only when the bandwidth of an individual band reaches $2\pi/T$. In contrast, the total bandwidth is irrelevant. This is fully consistent with Criterion~\ref{criterion_2}, and the model is reduced to the single-band case.

For comparison, we introduce a Type-II model defined by:
\begin{eqnarray}\label{eq:twoband2}
\hat{H}_{1(2)}^{(\text{II})}&&=t_1(\hat L+\hat R)\otimes[\hat \sigma_0+(1\pm i)\hat\sigma_x]\nonumber\\
&&+t_2(\hat L^2+\hat R^2)\otimes [\hat \sigma_0+(1\pm i)\hat\sigma_z],
\end{eqnarray}
with $\hat P$ being the identity matrix.
In this case, the non-commuting hopping matrices allow the non-Hermitian perturbation terms to mix the two bands when they approach each other.
The corresponding quasienergy spectrum is shown in Fig.~\ref{fig:twoband}(b). As predicted by Criterion~\ref{criterion_2}, unlike the Type-I case, EPs first emerge when the bandwidth of the total spectrum reaches $2\pi/T$. Specifically, the transition is triggered when the top of the upper band exceeds $\pi/T$ and intersects the bottom of the lower band.

\section{Dynamical signatures of $\mathcal{PT}$ breaking}
\label{sec:dynamics}

In this section, we examine the dynamical consequences of the
scale-free localization found in the $\mathcal{PT}$-broken phase.
We consider an initial state localized at the center of the chain,
\begin{equation}
    \ket{\psi(0)}=\ket{j_0},
    \qquad
    j_0= \lfloor(N+1)/2\rfloor,
\end{equation}
and evolve it stroboscopically according to
\begin{equation}
    \ket{\psi(nT)}=\hat U_F^n\ket{\psi(0)}.
\end{equation}
Here, $\lfloor\cdots\rfloor$ takes the integer part of a positive number. Because the Floquet evolution is generally nonunitary, all dynamical
observables are evaluated using the normalized state at each time.
We consider the normalized mean position
\begin{equation}\label{eq:meanposition}
    \frac{x_{\mathrm{avg}}(n)}{N}
    =
    \frac{1}{N}
    \frac{
        \bra{\psi(nT)}\hat X\ket{\psi(nT)}
    }{\bra{\psi(nT)}\ket{\psi(nT)}
    },
\end{equation}
where $ \hat X=\sum_{j=1}^{N}j\ket{j}\bra{j}$ is the position operator.

\begin{figure}[t]
    \hspace*{-0.5\textwidth}
    \includegraphics[width=\columnwidth]{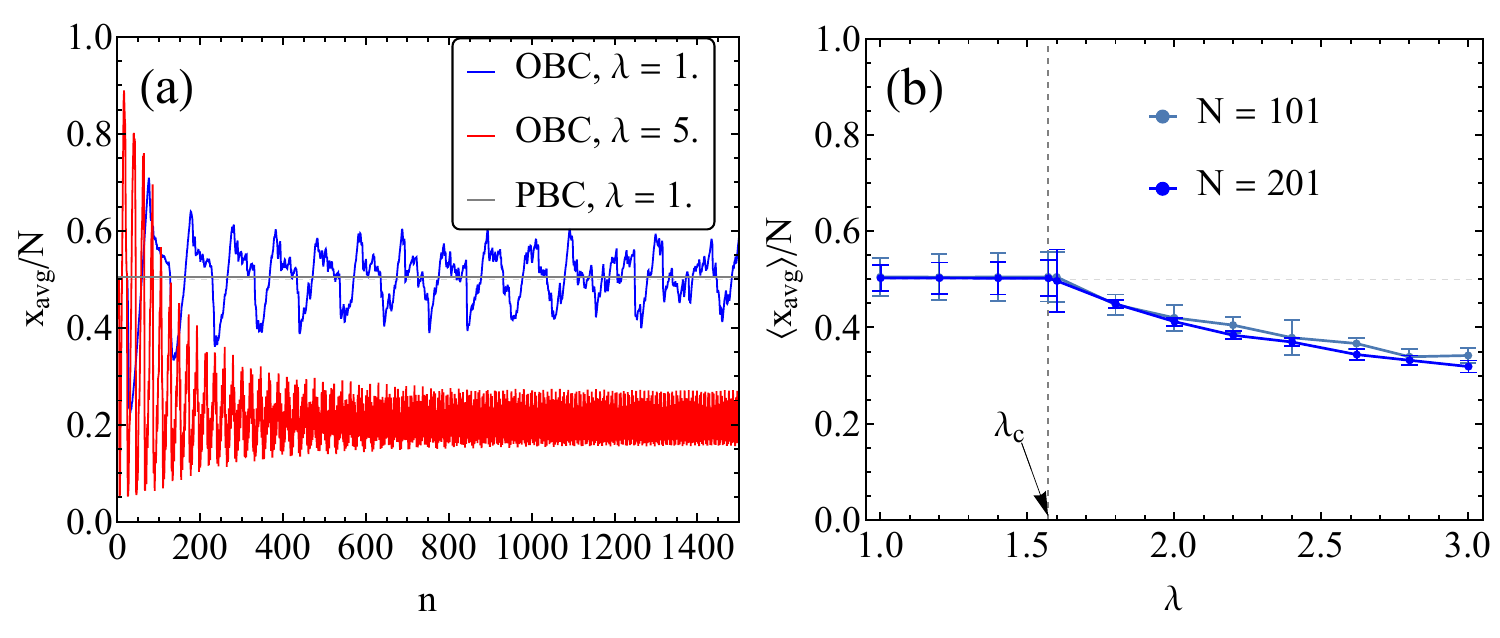}
    \caption{ Dynamical signature of scale-free localization in the toy model.
    (a) Normalized mean position $x_{\mathrm{avg}}/N$ as a function
    of the number of driving periods for an initial excitation
    localized at the center of the chain. Results are shown under PBC
    at $\lambda=1$ and under OBC in the $\mathcal{PT}$-unbroken
    ($\lambda=1$) and $\mathcal{PT}$-broken ($\lambda=5$) phases.
    (b) Late-time average $\overline{x}_{\mathrm{avg}}/N$ under OBC
    as a function of $\lambda$, evaluated over $n=3000$--$5000$ for $N=101$ and $n=8000$--$10000$ for $N=201$.
    The vertical dashed line marks the $\mathcal{PT}$-breaking
    threshold $\lambda_c=\pi/2$. The error bars exhibit temporal fluctuations within the averaging
    interval.
    }
    \label{fig:dynamics}
\end{figure}

Figure~\ref{fig:dynamics}(a) shows the stroboscopic evolution of
$x_{\mathrm{avg}}/N$ for the toy model [Eqs.~\eqref{eq:Hamiltonian} and~\eqref{eq:simple_model}]. Under PBC, the wave packet
remains centered around $x_{\mathrm{avg}}/N=1/2$. Under OBC in the
$\mathcal{PT}$-unbroken phase ($\lambda=1<\lambda_c$), the mean
position exhibits coherent oscillations but remains close to the
chain center on average. These oscillations originate from the
interference between different Floquet modes and repeated reflections
from the boundaries of the finite chain.

In the $\mathcal{PT}$-broken phase
($\lambda=5>\lambda_c$), by contrast, the dynamics develops a
pronounced bias toward one boundary. In this regime, Floquet modes
with positive imaginary quasienergies are exponentially amplified
relative to the decaying modes. The dominant growing modes possess
the boundary-asymmetric scale-free profiles shown in
Fig.~\ref{fig:scale_free}(b). Consequently, an initial state having a nonzero overlap
with these modes evolves into a normalized wave packet preferentially
distributed near one boundary. The persistent oscillations result
from interference between multiple growing modes with different real
quasienergies.

To extract the persistent boundary bias from these oscillations, we
define the late-time average
\begin{equation}
    \frac{\overline{x}_{\mathrm{avg}}}{N}
    =
    \frac{1}{n_2-n_1+1}
    \sum_{n=n_1}^{n_2}
    \frac{x_{\mathrm{avg}}(n)}{N},
    \label{eq:time_averaged_position}
\end{equation}
where $n_1=3000$ and $n_2=5000$. As shown in
Fig.~\ref{fig:dynamics}(b),
$\overline{x}_{\mathrm{avg}}/N$ remains close to $1/2$ below the
transition threshold $\lambda_c=\pi/2$, but deviates systematically
from the center after the system enters the $\mathcal{PT}$-broken
phase. Thus, the late-time boundary bias, rather than a monotonic
motion toward the boundary, provides a dynamical signature of the
boundary-asymmetric scale-free modes.

Here, we note that the observable is not always appropriate for characterizing scale-free localization. There are cases where the scale-free modes are symmetrically distributed towards the left and right boundaries, thereby eliminating the bias in $x_{\text{avg}}$. For these situations, an alternative quantity can be used to measure the wavefunction accumulation towards the boundaries; see details in Appendix~\ref{app:dynamics_twoband}.

\section{Discussion and conclusions}\label{sec:conclusion}

Based on the models presented above, a few remarks are in order. First, we emphasize that the presence of the NHSE in the driving Hamiltonians $H_{1,2}$ (only containing unidirectional hopping) is not necessary for our construction, as explicitly demonstrated by the model defined in Eq.~\eqref{eq:twoband2} in which each driving Hamiltonian individually respects spatial reflection symmetry (taking symmetric hopping) such that NHSE is prohibited. This constitutes a key distinction between our findings and the NHSE-induced sensitivity discussed by Hu \textit{et al.}~\cite{Hu2026Self}.
Second, our theoretical framework can be directly generalized to multi-step and even continuous-time driving protocols, requiring only minor modifications to the model constructions. 
Finally, we anticipate that our theoretical framework is accessible to current experimental platforms, including photonic quantum walk experiments~\cite{xiao2020non,PhysRevLett.126.230402} and synthetic photonic lattices~\cite{Ozdemir2019, El-Ganainy2018, doi:10.1126/science.aar7709}.

In summary, we identify a novel mechanism for $\mathcal{PT}$-symmetry breaking in one-dimensional Floquet systems. The non-Hermiticity in the effective Floquet Hamiltonian arises from the non-commutativity of the driving Hamiltonians within each period. We establish a set of general conditions ensuring that these non-Hermitian terms emerge only under OBC, while the system restores Hermiticity under PBC, which is unique to Floquet systems. We also unveil the fundamental relation between the $\mathcal{PT}$ symmetry breaking threshold and the winding of the quasienergy spectrum. Furthermore, the $\mathcal{PT}$-broken eigenmodes exhibit a unique scale-free localization, a feature that should facilitate the experimental detection of this novel phase transition.

\textit{Note added.---} We are aware of a related work by Hu \textit{et al.}~\cite{Hu2026Self}. 
That work introduces a driving protocol to engineer the scale-free localization by coupling pre-existing skin modes located at the frequency Brillouin zone boundaries $\pm\pi/T$. In contrast, we focus on the scale-free localization induced by boundary non-commutativity.

\section*{Acknowledgments}

We thank Zhong Wang for the support of this work at the early stage. We thank Yu-Min Hu for helpful discussions. B. L. is supported by the NSFC under Grant No.~12404185 and No.~12674203. H.-R.W. is supported by the NSFC (Grant No.~12125405 and No.~12342501) and the National Key R\&D Program of China (No.~2023YFA1406702). F. S. acknowledges support from NSFC under Grant No.~12404189 and from the Postdoctoral Fellowship Program of CPSF under Grant No.~GZB20240732.

\appendix

\section{Perturbation under OBC for model Eq.~\eqref{eq:simple_model}}\label{appendix:perturbation}
In this Appendix, we provide numerical evidence about the magnitude of perturbative matrix elements in the Floquet Hamiltonian.

\begin{figure}
\hspace*{-0.5\textwidth}
\begin{tabular}{cc}
 \includegraphics[width=1\linewidth]{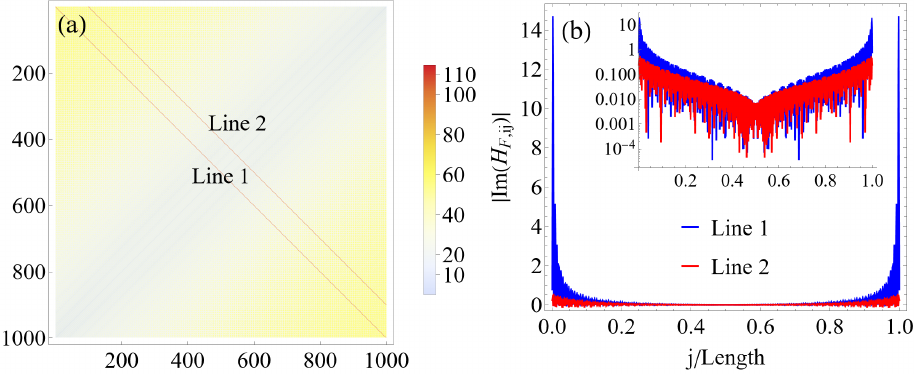}  \\
\includegraphics[width=1\linewidth]{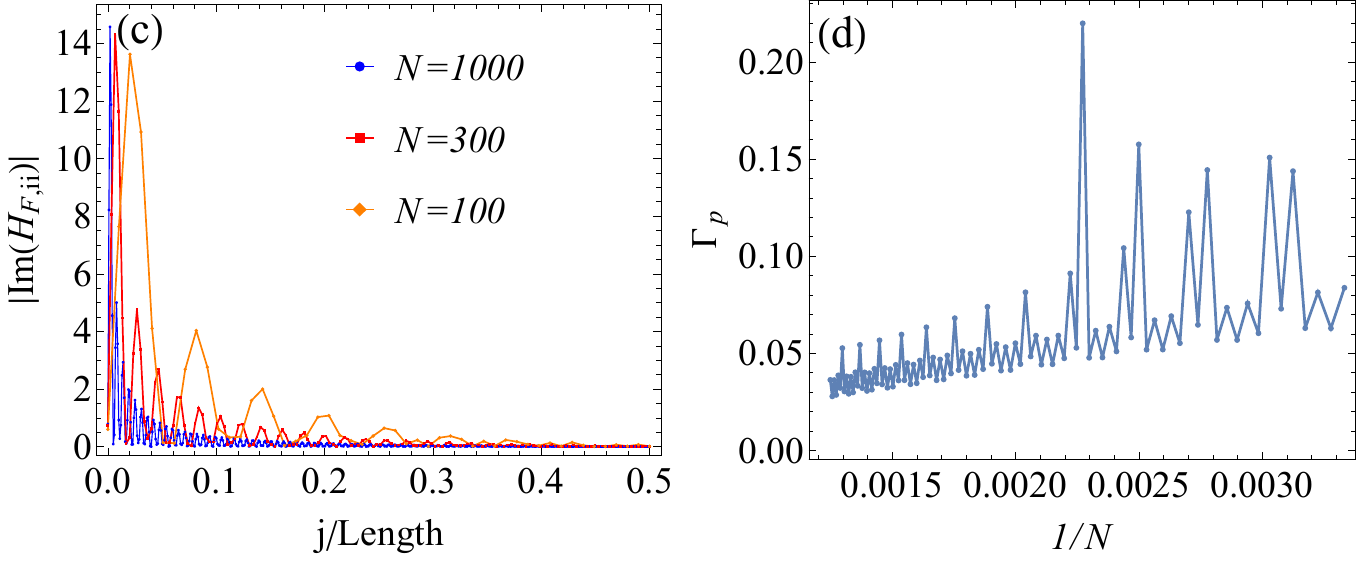} 
\end{tabular}
 \caption{(a) Magnitude of matrix elements of the non-Hermitian part of $H_{F,\text{OBC}}$, system size $N=1000$. (b) Diagonal and secondary diagonal matrix elements, as marked by solid lines in (a). The inset shows a logarithmically scaled plot. (c) Diagonal matrix elements for different system sizes $N$.  (d) Averaged bulk perturbation $\Gamma_p$ as a function of $1/N$. $\lambda=3$ for all the plots. }\label{fig:perturbation}
\end{figure}

In Figs.~\ref{fig:perturbation}(a) and (b), we plot the spatial profile of the non-Hermitian components of $H_{F,\text{OBC}}$ within the $\mathcal{PT}$-broken phase. These results clearly demonstrate that the non-Hermitian perturbations decay exponentially from the boundaries into the bulk.
In Fig.~\ref{fig:perturbation}(c), we compare the diagonal matrix elements across different system sizes $N$. It is evident that the localization of these boundary terms strengthens as the system size increases, even though the magnitude of the boundary elements remains essentially independent of $N$.

To quantify the system-size dependence of perturbations in the bulk region, we introduce the averaged bulk perturbation function:
\begin{eqnarray}
\Gamma_{p}(\lambda,N)=\frac{1}{(N-2s)^2}\sum_{i,j=s+1}^{N-s}|V_{ij}|,
\end{eqnarray}
where $V=H_{F,\text{OBC}}-H_0$, and $s$ is a cutoff parameter chosen to exclude sites near the boundaries. As shown in Fig.~\ref{fig:perturbation}(d), although $\Gamma_p$ exhibits significant oscillations for small system sizes, it asymptotically approaches $1/N$ scaling for sufficiently large $N$. This confirms that the non-Hermitian perturbations in the bulk are suppressed in the thermodynamic limit.

\section{Derivation of Eq.~\eqref{eq:real_quantity}}
\label{app:signature}

In this section, we provide the calculation details of the sign characteristic of the PBC Floquet
modes. We consider the two-step Floquet evolution Eq.~\eqref{eq:two_step_floquet_signature} with the involved Hamiltonians respecting Eq.~\eqref{eq:step_sharp_relation}. Therefore, evolution operator Eq.~\eqref{eq:two_step_floquet_signature} can be abbreviated as
\begin{equation}
U_F=(Y^\dagger)^{-1}Y.
\label{eq:UF_X_structure}
\end{equation}
where $Y\equiv e^{-iH_1T/2}$,
and the same construction applies both under PBC and OBC.

Under PBC, the eigenvalue equation of the evolution operator Eq.~\eqref{eq:UF_X_structure}, $U_{F}(k)|u_n(k)\rangle
=\xi_n(k)|u_n(k)\rangle$,  can be
rewritten as
\begin{equation}\label{eq:PBC_pencil}
Y(k)|u_n(k)\rangle
=
\xi_n(k)Y^\dagger(k)|u_n(k)\rangle.
\end{equation} 
Here, the eigenstates also satisfy
$\langle u_n(k)|U_F(k)=\xi_n(k)\langle u_n(k)|$,
because, by construction, the Floquet Hamiltonian
$H_F(k)=(i/T)\log U_F(k)$ is Hermitian under PBC,
and hence $U_F(k)$ is unitary.
Taking the expectation value of
Eq.~\eqref{eq:PBC_pencil} gives
\begin{align}
q_n(k)
&=
\xi_n(k)
\langle u_n(k)|
Y^\dagger(k)
|u_n(k)\rangle
\nonumber\\
&=
\xi_n(k)q_n^*(k).
\label{eq:qnk_constraint}
\end{align}
where $q_n(k)
=
\langle u_n(k)|
Y(k)|u_n(k)\rangle$.
For a $\mathcal{PT}$-unbroken PBC mode, $|\xi_n(k)|=1$, so we have the parameterization $\xi_n(k)=e^{-i\theta_n(k)}$.
Equation~\eqref{eq:qnk_constraint} then implies
\begin{equation}
e^{i\theta_n(k)/2}q_n(k)
=
\left[
e^{i\theta_n(k)/2}q_n(k)
\right]^*,
\end{equation}
implying this quantity is real.

\section{Derivation of the projected constraint
\(K^\dagger\eta=\eta K\)}
\label{app:projected_constraint}

In this Appendix, we derive the generalized-Hermiticity constraint obeyed
by the effective generator in a resonant Floquet subspace. The derivation
starts from the exact adjoint structure of the Floquet
operator Eq.~\eqref{eq:UF_X_structure}. No Hermiticity of the original driving Hamiltonians is assumed.

\subsection{Unperturbed resonant subspace}

From the evolution operator eigen equation $U_F|\psi\rangle=\xi|\psi\rangle$, and Eq.~\eqref{eq:UF_X_structure}, we obtain the generalized eigenvalue equation
\begin{equation}
\left(Y-\xi Y^\dagger\right)|\psi\rangle=0.
\label{eq:app_generalized_pencil}
\end{equation}
Consider two PBC modes that become resonant at the unit-circle eigenvalue 
\begin{equation}
\xi_0=e^{-i\theta_0},
\qquad
|\xi_0|=1.
\label{eq:app_z0}
\end{equation}
We denote the corresponding states by
\(\{|\psi_1\rangle,|\psi_2\rangle\}\). In the leading resonant problem,
they satisfy
\begin{equation}
\Omega_0|\psi_j\rangle=0,
\qquad
\Omega_0\equiv Y_0-\xi_0Y_0^\dagger,
\qquad
j=1,2,
\label{eq:app_A0}
\end{equation}
where \(Y_0\) denotes the unperturbed PBC operator at the resonance. Taking the Hermitian conjugate of \(\Omega_0\), we find
\begin{eqnarray}
\Omega_0^\dagger
&&=
Y_0^\dagger-\xi_0^*Y_0
\nonumber\\
&&=
-\xi_0^{-1}
\left(Y_0-\xi_0Y_0^\dagger \right)
\nonumber\\
&&=
-\xi_0^{-1}\Omega_0.
\label{eq:app_A0_adjoint}
\end{eqnarray}
Hence, taking the Hermitian conjugation of equation~\eqref{eq:app_A0} implies $\langle\psi_i|\Omega_0^\dagger=(-\xi_0^{-1})\langle\psi_i|\Omega_0=0$, leading to 
\begin{equation}
\langle\psi_i|\Omega_0=0.
\label{eq:app_left_null}
\end{equation}

\subsection{First-order projected Floquet problem}

The finite-size PBC detuning (compared to the exact momentum space result in thermodynamic limit) and the OBC boundary coupling both have
matrix elements of order \(N^{-1}\) between normalized extended states.
We therefore introduce $\epsilon=1/N$
and write, within the resonant subspace,
\begin{equation}
Y=Y_0+\epsilon\,\delta Y+O(\epsilon^2).
\label{eq:app_X_expansion}
\end{equation}
Here \(\delta Y\) includes both the finite-size detuning and the projected
boundary correction. Equation~\eqref{eq:app_X_expansion} is a projected
asymptotic expansion; it does not require the operator norm of the full
boundary perturbation to vanish as \(N^{-1}\). The Floquet eigenvalue near \(\xi_0\) is parameterized as
\begin{equation}
\xi
=
\xi_0e^{-i\epsilon\kappa}
=
z_0\left(1-i\epsilon\kappa\right)
+O(\epsilon^2).
\label{eq:app_z_expansion}
\end{equation}
The corresponding state is written as
\begin{equation}
|\Psi\rangle
=
\sum_{j=1}^{2}c_j|\psi_j\rangle
+
\epsilon|\delta\psi\rangle
+
O(\epsilon^2).
\label{eq:app_state_expansion}
\end{equation}

Substituting Eqs.~\eqref{eq:app_X_expansion}--\eqref{eq:app_state_expansion}
into Eq.~\eqref{eq:app_generalized_pencil}, we obtain
the first-order equation:
\begin{align}
A_0|\delta\psi\rangle
+
\left(
\delta Y
-\xi_0\delta Y^\dagger 
+i \xi_0\kappa Y_0^\dagger 
\right)
\sum_jc_j|\psi_j\rangle
=0.
\label{eq:app_first_order}
\end{align}
Multiplying Eq.~\eqref{eq:app_first_order} by
\(\langle\psi_i|\), the first term vanishes because of
Eq.~\eqref{eq:app_left_null}. Furthermore, Eq.~\eqref{eq:app_A0}
gives $\xi_0Y_0^\dagger |\psi_j\rangle
=
Y_0|\psi_j\rangle$.
We consequently obtain
\begin{equation}
\left(M+i\kappa Q\right)\bm c=0,
\label{eq:app_projected_equation}
\end{equation}
where \(\bm c=(c_1,c_2)^T\) and
\begin{eqnarray}
&&M_{ij}
=
e^{i\theta_0/2}
\langle\psi_i|
\left(
\delta Y-z_0\delta Y^\dagger 
\right)
|\psi_j\rangle,
\label{eq:app_M_definition}\\
&&Q_{ij}
=
e^{i\theta_0/2}
\langle\psi_i|Y_0|\psi_j\rangle
\label{eq:app_Q_definition}.
\end{eqnarray}
Eq.~\eqref{eq:app_projected_equation} is the projected Floquet eigen equation in the subspace spanned by two degenerate PBC modes. In the following, we will briefly discuss the properties of $M$ and $Q$ matrices.

\underline{\textbf{$M$ matrix}}. To uncover the properties of $M$, we define $B\equiv e^{i\theta_0/2}(\delta Y-\xi_0\delta Y^\dagger )$. Its Hermitian conjugate is
\begin{align}
B^\dagger
&=
e^{-i\theta_0/2}(\delta Y^\dagger-\xi_0^*\delta Y)
\nonumber\\
&=-e^{-i\theta_0/2}
\xi_0^{-1}
\left(
\delta Y-\xi_0\delta Y^\dagger 
\right)
\nonumber\\
&=
-B.
\label{eq:app_B_adjoint}
\end{align}
The operator \(B\) is thus anti-Hermitian. Its projection
onto the resonant subspace is also anti-Hermitian. Specifically, $M=P_{d}B P_{d}$, and hence
\begin{equation}
M^\dagger=P_d B^\dagger P_d=-M,
\label{eq:app_M_antiHermitian}
\end{equation}
where $P_d=\sum_{j=1,2}|\psi\rangle\langle \psi_j|$ is the projection operator, satisfying $P_d^\dagger=P_d$.

\underline{\textbf{$Q$ matrix}}.  From
Eq.~\eqref{eq:app_A0}, $Y_0^\dagger|\psi_j\rangle
=
\xi_0^{-1}Y_0|\psi_j\rangle$. It follows that
\begin{align}
Q_{ji}^*
&=
e^{-i\theta_0/2}
\langle\psi_i|Y_0^\dagger |\psi_j\rangle
\nonumber\\
&=
e^{-i\theta_0/2}\xi_0^{-1}
\langle\psi_i|Y_0|\psi_j\rangle
\nonumber\\
&=
e^{i\theta_0/2}
\langle\psi_i|Y_0|\psi_j\rangle
\nonumber\\
&=
Q_{ij},
\end{align}
proving that
\begin{equation}
Q^\dagger=Q,
\label{eq:app_Q_Hermitian}
\end{equation}
namely, $Q$ is a Hermitian matrix.

\subsection{Effective generator and the metric constraint}

Equation~\eqref{eq:app_projected_equation} can be written as
\begin{equation}
K\bm c
=
\kappa\bm c,
\end{equation}
where $K\equiv iQ^{-1}M$ is the effective generator.
The projected Floquet operator is correspondingly
\begin{equation}
U_{\mathrm{eff}}
=
\xi_0
\exp\left(-\frac{i}{N}K\right)
+
O(N^{-2}).
\label{eq:app_Ueff}
\end{equation}
Using \(Q^\dagger=Q\) and \(M^\dagger=-M\), we find
\begin{align}
K^\dagger Q
&=
\left(iQ^{-1}M\right)^\dagger Q
\nonumber\\
&=
-iM^\dagger
\left(Q^{-1}\right)^\dagger Q
\nonumber\\
&=
-iM^\dagger Q^{-1}Q
\nonumber\\
&=
iM.
\label{eq:app_KdaggerQ}
\end{align}
On the other hand,
\begin{equation}
QK
=
Q\left(iQ^{-1}M\right)
=
iM.
\label{eq:app_QK}
\end{equation}
Combining Eqs.~\eqref{eq:app_KdaggerQ} and \eqref{eq:app_QK}, we obtain
\begin{equation}
K^\dagger Q=QK.
\label{eq:app_KQ_constraint}
\end{equation}

Since \(Q\) is Hermitian and nonsingular before the EP, one can choose a basis transformation \(S\) such that
\begin{equation}
S^\dagger Q S
=
\eta
=
\begin{pmatrix}
\nu_1&0\\
0&\nu_2
\end{pmatrix},
\qquad
\nu_i=\pm1.
\label{eq:app_eta_definition}
\end{equation}
Defining $K_\eta=S^{-1}KS$,
Eq.~\eqref{eq:app_KQ_constraint} becomes $K_\eta^\dagger\eta=\eta K_\eta$. Dropping the subscript on \(K_\eta\), we arrive at
\begin{equation}
K^\dagger\eta=\eta K.
\label{eq:app_final_constraint}
\end{equation}
Thus, the effective generator is self-adjoint with respect to the signature
metric \(\eta\).

\section{Commutator under OBC}\label{appendix:commutator}
Here, we elaborate on the condition (iii) for general constructions presented in Eq.~\eqref{eq:general_form}. The commutator under OBC can be decomposed into two parts:
\begin{eqnarray}
[\hat H_1,\hat H_2]&&=\hat G_1+\hat G_2,
\end{eqnarray}
where
\begin{eqnarray}
\hat G_1&&=\mathbbm{1}\otimes[\hat A_1,\hat A_2]+\sum_{m=1}^w\hat L^m\otimes([\hat A_1, \hat X_2^{(m)}]-[\hat A_2, \hat X_1^{(m)}])\nonumber\\
&&+\hat R^m\otimes([\hat A_1, \hat Y_2^{(m)}]-[\hat A_2, \hat Y_1^{(m)}])\nonumber\\
&&+\sum_{m,n=1}^w\hat L^{m+n}\otimes[\hat X_1^{(m)},\hat X_2^{(n)}]+\hat R^{m+n}\otimes[\hat Y_1^{(m)},\hat Y_2^{(n)}]\nonumber\\
&&+\sum_{n=1}^w\sum_{m\leq n}\hat L^{n-m}\otimes([\hat X_1^{(n)}, \hat Y_2^{(m)}]+[\hat Y_1^{(m)}, \hat X_2^{(n)}])\nonumber\\
&&+\hat R^{n-m}\otimes([\hat Y_1^{(n)}, \hat X_2^{(m)}]+[\hat X_1^{(m)}, \hat Y_2^{(n)}]),
\end{eqnarray}
and
\begin{eqnarray}
\hat G_2&&=\sum_{n=1}^w\sum_{m\leq n}(\hat L^n\hat R^m-\hat L^{n-m})\otimes[\hat X_1^{(n)}, \hat Y_2^{(m)}]\nonumber\\
&&+(\hat R^n\hat L^m-\hat R^{n-m})\otimes[\hat Y_1^{(n)}, \hat X_2^{(m)}]\nonumber\\
&&+\sum_{n=1}^w\sum_{m< n}(\hat L^m\hat R^n-\hat R^{n-m})\otimes[\hat X_1^{(m)}, \hat Y_2^{(n)}]\nonumber\\
&&+(\hat R^m\hat L^n-\hat L^{n-m})\otimes[\hat Y_1^{(m)}, \hat X_2^{(n)}]\nonumber\\
&&+\sum_{m,n=1}^w[\hat L^m,\hat R^n]\otimes (\hat Y_2^{(n)}\hat X_1^{(m)}-\hat X_2^{(n)}\hat Y_1^{(m)}).\nonumber\\
\end{eqnarray}
The first part $\hat G_1$ represents the bulk contribution, and vanishes identically provided that condition (ii) is satisfied, i.e., the Bloch Hamiltonians commute: $[\hat h_1(k),\hat h_2(k)]=0$. In contrast, the second term $\hat G_2$ arises from boundary effects, and vanishes only when at least one of the inequalities in Eq.~\eqref{eq:cond3} is fulfilled.

\section{Decoupling of the two bands in the Type-I model}
\label{app:typeI}

In this Appendix, we show that the two bands of the Type-I
model in Eq.~\eqref{eq:twoband1} belong to two exactly decoupled internal
sectors, even in the presence of OBC-induced boundary
corrections. The two driving Hamiltonians can be written as
\begin{equation}
    \hat H_i^{(\mathrm{I})}
    =
    \hat Z_i\otimes\hat M,
    \qquad i=1,2,
    \label{eq:typeI_factorization}
\end{equation}
where $\hat Z_1=\frac{3}{2}\hat 1+\hat L, \hat Z_2=\frac{3}{2}\hat 1+\hat R, \hat M=t\hat\sigma_x+\hat\sigma_z$. The internal matrix $\hat M$ has eigenvalues $m_s=s\sqrt{1+t^2}$ ($s=\pm$)
and defines a conserved operator
\begin{equation}
    \hat Q
    =
    \hat 1\otimes
    \frac{\hat M}{\sqrt{1+t^2}}.
\end{equation}
Since $[\hat H_1^{(\mathrm{I})},\hat Q]
    =
    [\hat H_2^{(\mathrm{I})},\hat Q]
    =0$,  the internal label $s=\pm$ is conserved under both PBC and
OBC. In the eigenbasis of $\hat M$, the Floquet operator
therefore decomposes as
\begin{equation}
    \hat U_F
    =
    \hat U_F^{(+)}
    \oplus
    \hat U_F^{(-)},
\end{equation}
with
\begin{equation}
    \hat U_F^{(s)}
    =
    e^{-im_s\hat Z_2T/2}
    e^{-im_s\hat Z_1T/2}.
    \label{eq:typeI_sector_UF}
\end{equation}
The effective Floquet Hamiltonian defined on a fixed
quasienergy branch inherits the same block structure,
\begin{equation}
    \hat H_F
    =
    \hat H_F^{(+)}
    \oplus
    \hat H_F^{(-)},
    \qquad
    \hat H_F^{(s)}
    =
    \frac{i}{T}\log\hat U_F^{(s)}.
    \label{eq:typeI_sector_HF}
\end{equation}
Consequently, the OBC-induced correction
$\hat V=\hat H_{F,\mathrm{OBC}}-\hat H_0$ also commutes with
$\hat Q$. Its matrix elements between the two sectors vanish
identically:
\begin{equation}
    \bra{\psi_{+,a}}
    \hat V
    \ket{\psi_{-,b}}
    =0.
    \label{eq:typeI_selection_rule}
\end{equation}

The absence of inter-sector coupling is also evident from the
leading BCH correction:
\begin{align}
    [\hat H_1^{(\mathrm{I})},\hat H_2^{(\mathrm{I})}]
    &=
    [\hat z_1,\hat z_2]\otimes\hat M^2
    \nonumber\\
    &=
    (1+t^2)[\hat L,\hat R]\otimes\hat\sigma_0.
    \label{eq:typeI_commutator}
\end{align}
Under OBC, $[\hat L,\hat R]$ generates boundary-supported
terms, but their internal component is proportional to
$\hat\sigma_0$ and hence cannot mix the two sectors. More
generally, all higher-order nested commutators contain only
powers of $\hat M$ and therefore remain diagonal in its
eigenbasis.

It follows that a quasienergy crossing between the $s=+$ and
$s=-$ sectors is an ordinary symmetry-protected degeneracy,
rather than an exceptional point. Exceptional points can
emerge only within an individual sector, where the
OBC-induced correction acts nontrivially. The single-band
PT-breaking condition must therefore be applied separately
to each sector: the relevant threshold is determined by the
bandwidth of an individual sector, whereas the total
bandwidth of the two decoupled sectors is irrelevant.

\section{Dynamical signatures of scale-free localization in two-band models}
\label{app:dynamics_twoband}

\begin{figure}[t]
    \hspace*{-0.5\textwidth}
    \includegraphics[width=\columnwidth]{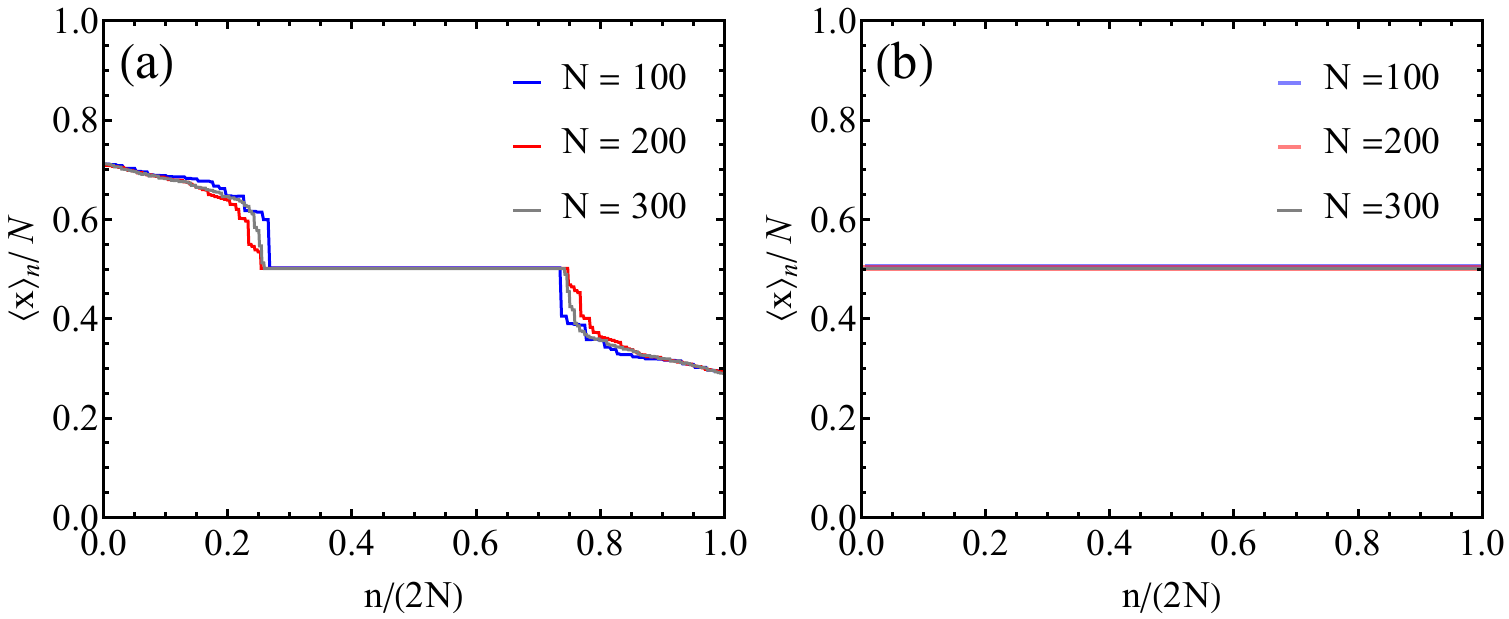}
    \caption{The eigenstate-resolved mean position ratio $\langle x\rangle_n/N$ for two-band models, with (a) and (b) respectively for type-I [Eq.~\eqref{eq:twoband1}] and type-II [Eq.~\eqref{eq:twoband2}] models. The curves in each panel are plotted for various system sizes $N$. The eigenstate index $n$ is sorted in ascending order of the imaginary part of the corresponding quasienergy. }
    \label{fig:eigenstate_center_twoband}
\end{figure}

In this section, we extend the discussion on dynamical signatures to the two-band models in Sec.~\ref{sec:multi-band}. To start with, we first explore the distribution of wavefunction profiles for two-band models by considering the eigenstate-resolved mean position $\langle x\rangle_n/N$ introduced in the main text. Fig.~\ref{fig:eigenstate_center_twoband} shows that the $\mathcal{PT}$-broken eigenstates for type-I and type-II models are distributed asymmetrically and antisymmetrically with respect to the center of the systems. For the case with symmetric distribution, the dynamics of the mean position in Eq.~\eqref{eq:meanposition} is not a good indicator for the scale-free localization anymore. Therefore, a new indicator is needed to characterize the scale-free localization in such a situation. In the following, we will separately study the dynamical features for the type-I and type-II models.

\begin{figure}[t]
    \hspace*{-0.5\textwidth}
    \includegraphics[width=\columnwidth]{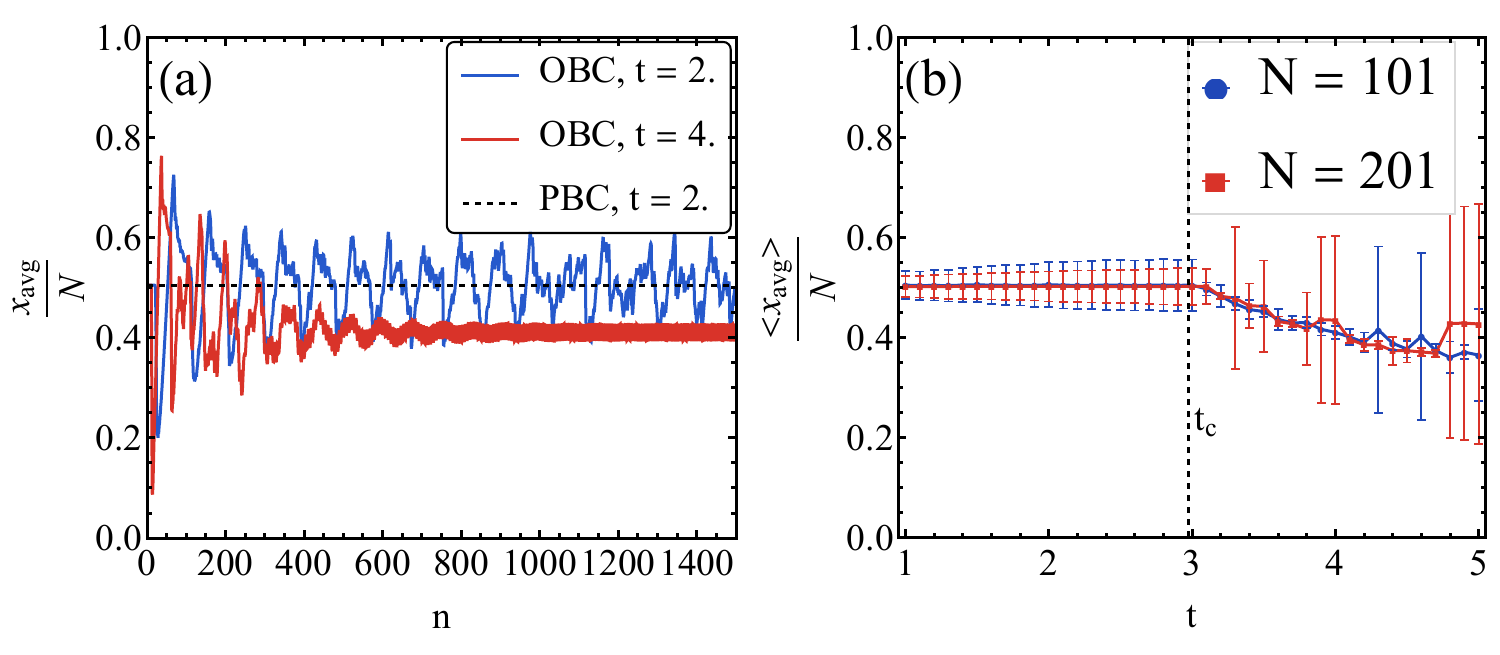}
    \caption{ Dynamical signature of scale-free localization in the Type-II
    model.
    (a) Normalized mean distance from the chain center,
    $D(n)=2\langle|x-x_c|\rangle/(N-1)$, for an initially localized
    state. Results are shown under PBC and under OBC on the two sides
    of the $\mathcal{PT}$-breaking transition.
    (b) Late-time average $\overline{D}$ under OBC as a function of
    $t_1$, evaluated over $n=4000-7000$ for $N=101$ and $n=8000-10000$ for $N=201$. The horizontal dashed line denotes the extended-state
    reference value $D=1/2$ in the thermodynamic limit, and the
    vertical dashed line marks the $\mathcal{PT}$-breaking threshold.
    The error bars represent temporal fluctuations within the
    averaging interval.
    }
    \label{fig:typeI_dynamics}
\end{figure}

For the type-I model, the PT-broken eigenstate distributes towards a boundary, similar to the toy model in the main text. Therefore, we can still use  $x_{\text{avg}}(n)$ to detect the feature of scale-free localization. For a two-band model, we will assume the initial state equally occupies the central unit cell.  As shown in Fig.~\ref{fig:typeI_dynamics}, in the $\mathcal{PT}$-broken phase, the averaged position evolves to a value deviating from the center of the system, and the deviation of the late-time mean position clearly indicates the $\mathcal{PT}$-breaking point. 

The mean position is particularly effective for the toy model because
its growing scale-free modes are asymmetric between the two
boundaries. It is, however, insensitive to a reflection-symmetric
redistribution of the wave packet. In particular, simultaneous
accumulation near both boundaries can still give
$x_{\mathrm{avg}}/N\simeq1/2$. To characterize such dynamics in the
Type-II model, we therefore introduce the normalized mean distance
from the center,
\begin{equation}
    D(n)
    =
    \frac{2}{N-1}
    \frac{
        \bra{\psi(nT)}
        \left|\hat X-x_c\right|
        \ket{\psi(nT)}
    }{
        \braket{\psi(nT)}
    },
    \label{eq:mean_center_distance}
\end{equation}
with $x_c=\frac{N+1}{2}$.
Here,
\begin{equation}
    \left|\hat X-x_c\right|
    =
    \sum_{j=1}^{N}
    \left|j-x_c\right|
    \ket{j}\bra{j},
\end{equation}
with an identity operator in the internal two-band space understood.
The normalization is chosen such that $D=0$ for a state localized at
the center and $D=1$ for a state localized at either boundary. A
spatially uniform distribution gives
\begin{equation}
    D_{\mathrm{uni}}
    =
    \frac{N}{2(N-1)}
    \longrightarrow \frac{1}{2}
\end{equation}
for even $N$ in the thermodynamic limit. Therefore, $D>1/2$
indicates that the wave-function weight is biased toward the
boundaries, irrespective of whether the accumulation occurs at one
or both ends.

\begin{figure}[t]
    \hspace*{-0.5\textwidth}
    \includegraphics[width=\columnwidth]{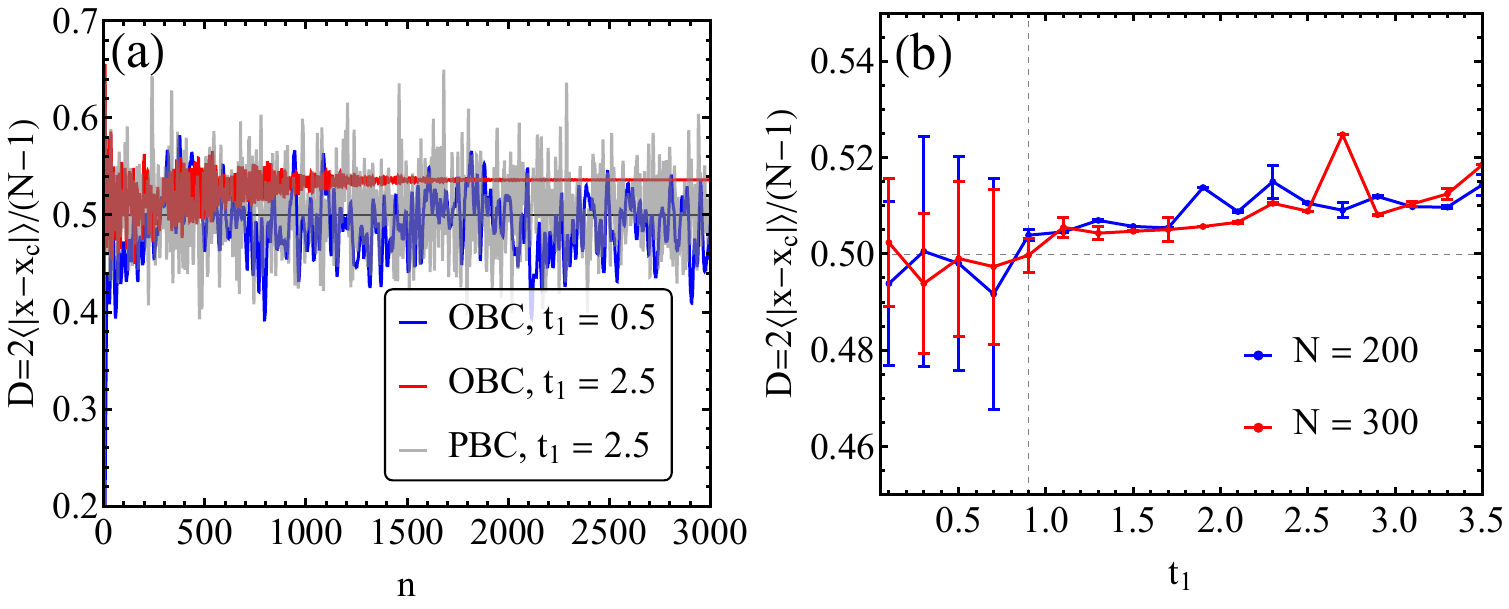}
    \caption{ Dynamical signature of scale-free localization in the Type-II
    model.
    (a) Normalized mean distance from the chain center,
    $D(n)=2\langle|x-x_c|\rangle/(N-1)$, for an initially localized
    state. Results are shown under PBC and under OBC on the two sides
    of the $\mathcal{PT}$-breaking transition.
    (b) Late-time average $\overline{D}$ under OBC as a function of
    $t_1$, evaluated over $n=7000-10000$ for $N=200$ and $n=10000-15000$ for $N=300$. The horizontal dashed line denotes the extended-state
    reference value $D=1/2$ in the thermodynamic limit, and the
    vertical dashed line marks the $\mathcal{PT}$-breaking threshold.
    The error bars represent temporal fluctuations within the
    averaging interval.
    }
    \label{fig:typeII_dynamics}
\end{figure}

Figure~\ref{fig:typeII_dynamics}(a) shows the evolution of $D(n)$ for
the Type-II model. Under PBC, $D(n)$ fluctuates around the extended-state
reference value $1/2$. A similar behavior is found under OBC in the
$\mathcal{PT}$-unbroken phase, where the fluctuations reflect coherent
propagation and boundary reflections but do not produce a persistent
outward redistribution. In the $\mathcal{PT}$-broken phase, however,
$D(n)$ approaches a value systematically above $1/2$. The wave packet
therefore acquires enhanced weight away from the center and toward the
two boundaries, even though its mean position may remain close to
$x_{\mathrm{avg}}/N=1/2$ because of reflection symmetry.

We further define the late-time average
\begin{equation}
    \overline{D}
    =
    \frac{1}{n_2-n_1+1}
    \sum_{n=n_1}^{n_2}D(n).
    \label{eq:time_averaged_distance}
\end{equation}
As shown in Fig.~\ref{fig:typeII_dynamics}(b), for $n_1=4000$ and $n_2=7000$,
$\overline{D}$ remains close to the extended-state reference value
below the $\mathcal{PT}$-breaking threshold, whereas it becomes
systematically larger after the transition. The error bars represent
the temporal fluctuations of $D(n)$ within the averaging interval.
The onset of $\overline{D}>1/2$ is consistent with the appearance of
growing scale-free modes whose probability weight is enhanced near
the boundaries.

The two dynamical observables therefore probe complementary aspects
of scale-free localization. The mean position
$x_{\mathrm{avg}}/N$ detects a left--right asymmetry of the dominant
modes, while the mean center distance $D$ detects boundary
accumulation even when the two boundaries are populated symmetrically.
Together, Figs.~\ref{fig:dynamics},~\ref{fig:typeI_dynamics}, and
\ref{fig:typeII_dynamics} demonstrate that the scale-free Floquet
modes leave observable dynamical signatures after the
$\mathcal{PT}$-breaking transition.

\bibliography{Floquet,dirac}

\end{document}